\documentclass[aps,prl,twocolumn]{revtex4}
\usepackage{amsfonts, amssymb, amsmath,graphicx}
\usepackage{subfigure}
\usepackage{multirow}
\usepackage{dcolumn}
\usepackage{bm}
\usepackage{color}

\begin{document}

\title{Synthetic Dimensions with Magnetic Fields and Local Interactions in Photonic Lattices}

\author{Tomoki Ozawa}
\author{Iacopo Carusotto}
\affiliation{INO-CNR BEC Center and Dipartimento di Fisica, Universit\`a di Trento, I-38123 Povo, Italy}%

\date{\today}

\newcommand{\iac}[1]{{\color{red} #1}}
\newcommand{\tom}[1]{{\color{blue} #1}}

\begin{abstract}
We discuss how one can realize a photonic device that combines synthetic dimensions and synthetic magnetic fields with spatially local interactions. Using an array of ring cavities, the angular coordinate around each cavity spans the synthetic dimension. The synthetic magnetic field arises as the intercavity photon hopping is associated with a change of angular momentum. Photon-photon interactions are local in the periodic angular coordinate around each cavity. Experimentally observable consequences of the synthetic magnetic field and of the local interactions are pointed out.
\end{abstract}

\maketitle


Dimensionality plays a key role in modern physics~\cite{ChaikinLubensky}. From the perspective of condensed matter physics, systems of dimensions 4 or higher were long considered relatively featureless because their properties are typically well captured by mean-field theories. This expectation was overturned by recent developments in the study of topological phases of matter, which hinted at rich novel physics in higher dimensional systems~\cite{Chiu:2016}.  In particular, analogues of the quantum Hall effect are predicted for any even dimensions, with new topological invariants appearing from four dimensions~\cite{Zhang:2001}.

While such predictions are clearly unaccessible in traditional condensed matter systems, a novel approach of simulating higher-dimensional topological models using ``synthetic dimensions" has very recently moved its first steps. The idea was first proposed in ultracold atomic gases, where the discrete internal spin degrees of freedom of the atoms are regarded as an extra synthetic dimension along which tunneling occurs via suitably designed Raman transitions~\cite{Boada:2012, Celi:2014,Mancini:2015,Stuhl:2015}. Synthetic dimensions have also been considered for other systems such as different electronic states of an atom~\cite{Livi:2016}, atoms in a harmonic potential~\cite{POG}, or light in arrays of multimode cavities~\cite{Schmidt:2015, Luo:2015, Ozawa:2016, Yuan:2016}.

Such developments in synthetic dimensions open a prospect of exploring physics in four dimensions using physically three-dimensional lattices. As first steps, proposals to observe the quantized magnetoelectric conductance of the four-dimensional integer quantum Hall effect have been put forward~\cite{Ozawa:2016, Price:2015, Price:2016}. Going beyond to explore fractional quantum Hall states is however facing great difficulties because of the very long range nature of interactions along the synthetic dimension in both atomic and photonic systems~\cite{Barbarino:2015, Zeng2015, Zoller:2015}. It is therefore of great interest to find schemes to realize synthetic dimensions where interactions are short ranged.

\begin{figure}[htbp]
\begin{center}
\includegraphics[width=8.5cm]{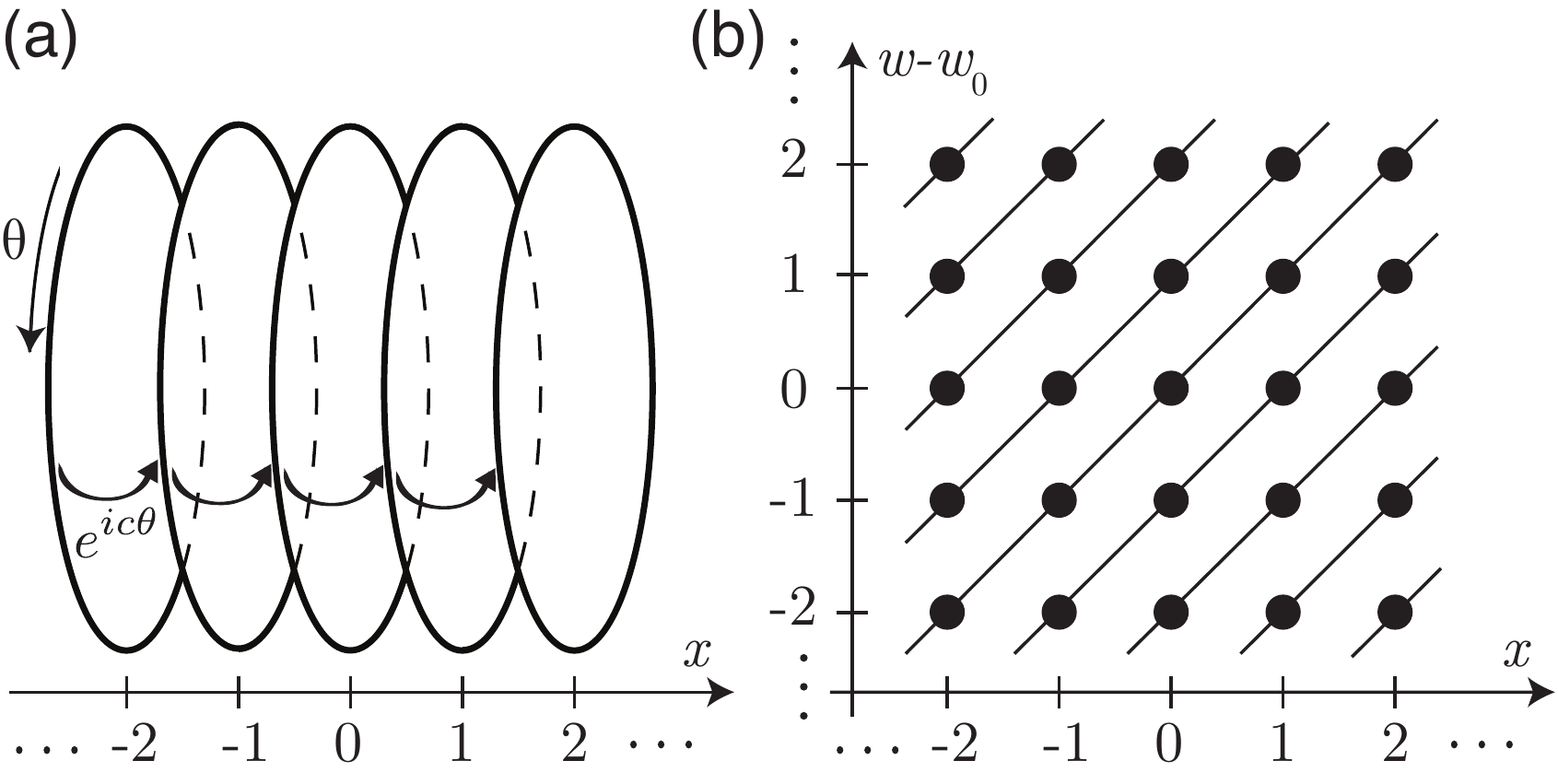}
\caption{Schematic illustration of hoppings (a) in the $x$-$\theta$ plane and (b) in the $x$-$w$ plane when $c = 1$.}
\label{schematic}
\end{center}
\end{figure}

To overcome this difficulty, we propose a general framework to realize a photonic lattice that combines synthetic dimensions and synthetic magnetic fields with local interactions. Our proposal therefore has the potential of hosting strongly correlated topologically nontrivial states of matter. We consider an array of ring-shaped photonic cavities. Differently from previous proposals~\cite{Luo:2015, Ozawa:2016, Yuan:2016} where the focus was on the mode index $w$, the synthetic dimension is here spanned by the geometrical angular coordinate $\theta$ around the disk that is conjugate to $w$. As higher dimensional cases do not display any conceptual difference, we will present our idea on the intuitively most transparent case of a one-dimensional chain of cavities along the $x$ direction. In this way, an effectively two-dimensional model in the $x$-$\theta$ plane is obtained, and a synthetic magnetic field naturally appears if the resonators are designed in a way that hopping along the $x$ direction is accompanied by a change in $w$. The spatially local photon-photon interactions within a cavity lead to the desired local interaction along both $x$ and $\theta$. We confirm the validity of our framework by numerical simulations of the cyclotron motion and the expansion dynamics of a suitably initialized wave packet. 

\textit{The model Hamiltonian}.---
Our target is to realize a Hamiltonian of the form
\begin{align}
	&\mathcal{H}
	=
	\sum_x \int_0^{2\pi}d\theta
	\left[
	\frac{D}{2}
	\{i\nabla_\theta b^\dagger_x (\theta)\}
	\{-i\nabla_\theta b_x (\theta)\}
	\right.
	\label{effham}\\
	&\left.-J \left\{ e^{ic\theta}b^\dagger_{x+1}(\theta) b_x(\theta) + \mathrm{H.c.}\right\}
	+ \frac{U}{2} b^\dagger_x(\theta) b^\dagger_x(\theta) b_x(\theta) b_x(\theta)
	\right], \notag
\end{align}
where $b_x (\theta)$ is an annihilation operator of a photon in a ring cavity at site $x$ with $\theta$ being the angular variable around a ring cavity. The site index $x$ is discrete, whereas the angular variable $\theta$ is continuous, taking a value $0 \le \theta < 2\pi$ with periodic boundary conditions $b_x(2\pi)=b_x(0)$.
The first term in Eq.~(\ref{effham}) is the kinetic energy term with an effective mass $1/D$ along the $\theta$ direction.  The second term describes hoppings between adjacent cavities of (real) amplitude $J$ and a $\theta$-dependent hopping phase equal to $c\theta$ as one hops from sites $(x,\theta)$ to $(x+1,\theta)$. The hopping phase represents an effective magnetic field of the strength $B_\mathrm{eff} = -c$ in the $x$-$\theta$ plane. Because of the periodicity in the $\theta$ direction, the constant $c$ must be an integer. This Hamiltonian describes the motion of a particle on a cylinder with a tight periodic potential along the axial direction, as schematically described in Fig.~\ref{schematic}(a). Spatially local photon-photon interactions in the $x$-$\theta$ plane are described by the last term.

In order to see how the Hamiltonian (\ref{effham}) can be physically realized, we Fourier transform the $b_x(\theta)$ operators to the conjugate angular momentum $w$ space:
\begin{align}
	b_{x,w}
	\equiv
	\sqrt{\frac{1}{2\pi}} \int_0^{2\pi} d\theta\, b_x(\theta) e^{-i(w-w_0)\theta}; \label{fouriertheta}
\end{align}
the reference angular momentum $w_0$ is introduced here for later convenience. In terms of the transformed $b_{x,w}$ operators, the Hamiltonian (\ref{effham}) is rewritten in the form
\begin{multline}
	\mathcal{H}=
	\\
	\sum_{x,w}\frac{D}{2}(w-w_0)^2 b^\dagger_{x,w}b_{x,w}
	-J
	\sum_{x,w} \left( b^\dagger_{x+1,w+c} b_{x,w} + \mathrm{H.c.} \right)
	\\
	+
	\frac{U}{4\pi}\sum_{x}\sum_{w_1 +w_2 = w_3 + w_4}
	b^\dagger_{x,w_1} b^\dagger_{x,w_2} b_{x,w_3} b_{x,w_4}. \label{hamiltonian}
\end{multline}
The hopping between adjacent cavities, described by the second term, is accompanied by a change of $c$ units of angular momentum $w$.
These hopping processes in the $x$-$w$ plane are schematically described in Fig.~\ref{schematic}(b). The last term represents the interphoton interaction which conserves the total angular momentum. 

\textit{Physical implementation}.---
While our proposal can be applied to a variety of photonic systems where photons follow a circular path inside a cavity, such as exciton-polariton micropillars~\cite{Deveaud:2016}, circuit QED setups~\cite{Koch:2010}, or surface plasmons~\cite{Gao:2016}, in the following we focus on a specific implementation scheme based on an array of silicon microring resonators~\cite{Hafezi:2013}.

We propose to design microring resonator cavities in a way that their mode frequencies depend on $x$ and $w$ as
\begin{align}
	\Omega_{x,w}
	=
	\Omega_{0} + \Omega_\mathrm{FSR} (w - w_0 - xc) + D(w-w_0)^2/2,
	\label{Omega_xw}
\end{align}
where $\Omega_0$ is the frequency of a reference mode $w_0\gg 1$ in the central $x=0$ cavity and we restrict our attention to $w>0$ modes. The free spectral range of the cavities is $\Omega_\mathrm{FSR}$ and the quadratic dispersion term $D$ accounts for deviations from perfect mode equispacing. In particular, the modes of neighboring cavities are assumed to be shifted by $-c\Omega_\mathrm{FSR}$, so that $\Omega_{x,w}\approx\Omega_{x+1,w+c}$.
The condition (\ref{Omega_xw}) can be obtained by careful fabrication of cavities adjusting available parameters such as the radius and the shape of the section of cavities~\cite{FernandoPC}.

If the hopping is much smaller than the free spectral range $J\ll \Omega_\mathrm{FSR}$, since $\Omega_{x,w}\approx\Omega_{x+1,w+c}$, the only effective tunneling processes are the ones associated with an angular momentum change by $c$.
Assuming that the relevant amplitude does not significantly depend on $x$ or on $w$, the noninteracting tight-binding Hamiltonian is
\begin{align}
	\tilde{\mathcal{H}}_0 =
	\sum_{x,w} \Omega_{x,w}a^\dagger_{x,w} a_{x,w}
	-
	J \sum_{x,w} \left( a^\dagger_{x+1, w+c} a_{x,w} + \mathrm{H.c.}\right),
\end{align}
where $a_{x,w}$ is the annihilation operator of a photon with the angular momentum $w$ at site $x$.
We assume for simplicity $J > 0$, but our results are robust with respect to the $w$-dependent hopping phases that appear in some implementations (see the Supplemental Material for further discussion~\cite{SM}).
Moving to a rotating frame by
$b_{x,w} \equiv a_{x,w} \exp\{i [ \Omega_0 + \Omega_\mathrm{FSR} (w-w_0-xc)] t\}$,
one recovers the noninteracting part of the desired Hamiltonian (\ref{hamiltonian}). Transformation to the $b_{x,w}$ operators corresponds to moving to a frame that rotates at $\Omega_{\mathrm{FSR}}$. The origin of the synthetic magnetic field can be understood by noting that in terms of the rescaled $\tilde{w} \equiv w - w_0 - xc$, which does not change during tunneling, the mode frequency (\ref{Omega_xw}) recovers that of a free particle in a magnetic field with the Landau gauge with a boost in the synthetic direction, $\Omega_{x,\tilde{w}}=\Omega_{0} + \Omega_\mathrm{FSR} \tilde{w} + \frac{D}{2}(\tilde{w} + xc)^2$, where $\tilde{w}$ is analogous to momentum in the synthetic dimension.

The photons in a cavity can interact via the nonlinearity of the underlying medium.
The electric field $\mathbf{E}(\mathbf{r},t)$ in a single cavity can be written as
\begin{align}
	\mathbf{E}(\mathbf{r},t)
	=
	\sum_w \left[ \mathcal{E}_w (\mathbf{r}) a_{w} + \mathcal{E}_w^* (\mathbf{r}) a^\dagger_{w} \right], \label{electric}
\end{align}
where, assuming a circular form for the cavity and restricting to radially polarized modes, the electric index profile $\mathcal{E}^w (\mathbf{r})$ of the mode $w$ in the cylindrical coordinates $(r,\phi,z)$ has the form
$	\mathcal{E}^w (\mathbf{r}) = R_w (r) Z_w (z) e^{iw \phi} \hat{e}_r$
where $\hat{e}_r$ is the unit vector in the radial direction.
Assuming that the $\chi^{(3)}$ nonlinearity is local in space and sufficiently fast in time, the photon-photon interactions are described by the Hamiltonian
\begin{align}
	\mathcal{H}_{\mathrm{int}}
	=
	\tilde{U} \int d^3 r \left| \mathbf{E}(\mathbf{r},t) \right|^4. \label{chi3}
\end{align}
Inserting Eq.~(\ref{electric}) into Eq.~(\ref{chi3}), one notices that only angular momentum conserving terms survive after the integration over $\mathbf{r}$. Under the rotating wave approximation, we ignore terms involving unequal numbers of creation and annihilation operators. 
Grouping all numerical factors into the coefficient $U$, one obtains the following interaction Hamiltonian in the rotating basis
\begin{align}
	\mathcal{H}_{\mathrm{int}}
	=
	\frac{U}{4\pi}\sum_{w_1 + w_2 = w_3 + w_4}
	b^\dagger_{w_1} b^\dagger_{w_2} b_{w_3} b_{w_4}, \label{interaction0}
\end{align}
which is straightforwardly extended to the many-cavity case described in our target Hamiltonian (\ref{hamiltonian}). 

The local interaction in $\theta$ requires that the nonlinearity be sufficiently fast on the time scale of $\Omega_\mathrm{FSR}$.
If the nonlinearity is slow, each term in Eq.~(\ref{interaction0}) is multiplied by a coefficient that is a function of $\Omega_\mathrm{FSR}(w_1-w_4)$, which decays on the scale of the inverse response time of the medium.
This condition can be physically understood by noting that a photon wave packet quickly rotates around the ring cavity at an angular speed $\Omega_\mathrm{FSR}$.

\textit{Single-particle physics}.---
We now turn our attention to the single-particle (noninteracting) physics predicted by the model Hamiltonian (\ref{effham}). This Hamiltonian describes a quantum particle subject to a magnetic field and moving in a hybrid geometry with one discrete and one continuous dimension. A charged particle in a magnetic field with two continuous dimensions is a text-book problem leading to the highly degenerate Landau levels~\cite{Landau:Book}. On the other hand, the same problem in two discrete dimensions is known as the Harper-Hofstadter model and leads to the energy spectrum known as the Hofstadter butterfly~\cite{Hofstadter:1976}. The configuration considered here lies in between the two cases of Landau levels and the Hofstadter butterfly and is closely related to Kane {\it et al.}'s coupled quantum wire construction~\cite{Kane:2002}.

When the system is periodic in the $x$ direction, the Hamiltonian (\ref{effham}) is translationally invariant in the $x$ direction, so one can take advantage of the conserved momentum $k_x$ and express
$
	b_x(\theta) = \frac{1}{\sqrt{N_x}} \sum_{k_x} e^{ik_x x} b(k_x, \theta),
$
where $N_x$ is the number of cavities along $x$. The problem is then reduced to the one-dimensional Hamiltonian
\begin{align}
	\mathcal{H}_0
	=
	\sum_{k_x} \int_0^{2\pi} &d\theta 
	\left[
	\frac{D}{2}
	(i\nabla_\theta b^\dagger (k_x, \theta))
	(-i\nabla_\theta b_x (k_x, \theta))
	\right.
	\notag \\
	&\left.
	-2J \cos \left( c\theta - k_x \right) b^\dagger (k_x, \theta) b (k_x, \theta)
	\right] \label{diagonalkx}
\end{align}
describing the mass of $1/D$ particles moving in a cosine-shaped potential with periodic boundary conditions $\theta + 2\pi = \theta$.
As the $k_x$ dependence in the potential $\cos (c\theta - k_x)$ can be absorbed into the shift of the origin of the periodic coordinate $\theta$, the one-particle spectrum does not depend on $k_x$. This implies that each state is $N_x$-fold degenerate, which is reminiscent of the Landau levels.

\begin{figure}[htbp]
\begin{center}
\includegraphics[width=8.5cm]{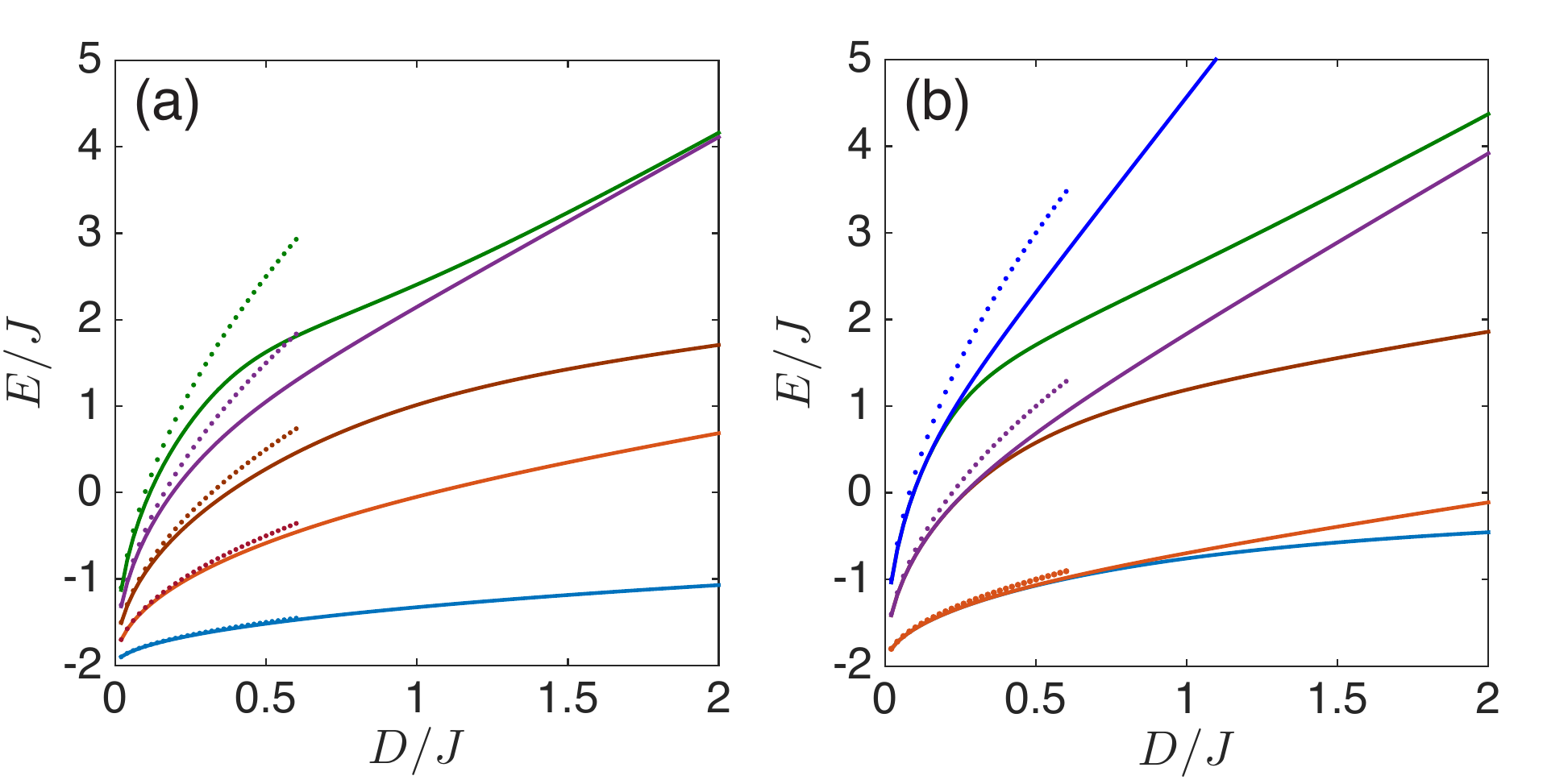}
\caption{Energy levels as a function of $D/J$ for (a) $c=1$ and (b) $c=2$. The solid lines show the numerically obtained first (a) five and (b) six energy eigenvalues, and the dotted lines are the analytical approximations for $D/J\ll 1$.}
\label{spectrum}
\end{center}
\end{figure}

In the limit of large $D$, the harmonic trapping in the $w$ direction dominates over the hopping along $x$, and the spectrum becomes $n^2 D/2$, where $n$ is an integer. In the opposite $D \ll J$ limit, when $c \neq 0$, the cosine potential $-2J \cos (c\theta)$ is large compared to the kinetic energy; therefore the low-lying energy levels of the system can be obtained with a quadratic approximation of the cosine potential around each of its $|c|$ minima. As in a harmonic oscillator, the energy levels are then approximately equispaced,
$
	E_n = -2J + \sqrt{2D/J}\left( n + 1/2 \right)cJ,
$
where $n \ge 0$ is a non-negative integer and, for a given value of $k_x$, each of them is approximately $|c|$-fold degenerate. The splitting due to tunneling between minima of the cosine potential is exponentially small in the $D\to 0$ limit. 
In Fig.~\ref{spectrum}, we plot the first several single particle energy levels as one varies $D$ for $c = 1$ and $2$: a very good agreement is found at small $D$ between the asymptotic analytical expression (dotted lines) and the full spectrum (solid lines).

Because of its intrinsic periodicity in $\theta$, our system does not have any edge in $\theta$. Open boundary conditions can be chosen along $x$, which lead to chiral edge states, whose properties are similar to the Landau level problem; more details are discussed in the Supplemental Material~\cite{SM}. In what follows, we focus our attention on experimentally observable magnetic and interaction effects in the bulk.

\textit{Cyclotron orbits}.---
As a first example, we investigate the semiclassical cyclotron orbits of a Gaussian wave packet in the $x$-$\theta$ plane. Such a wave packet can be created by a Gaussian pulse in the $x$ direction that overlaps with many $w$ modes of each cavity. This can be obtained if each cavity is locally coupled to the waveguide where the pulse is launched, so that the temporal shape of a short pulse of duration $\tau_P\ll \Omega_\mathrm{FSR}^{-1}$ gets transferred to the initial profile of the wave packet along $\theta$.

\begin{figure}[htbp]
\begin{center}
\includegraphics[width=8.0cm]{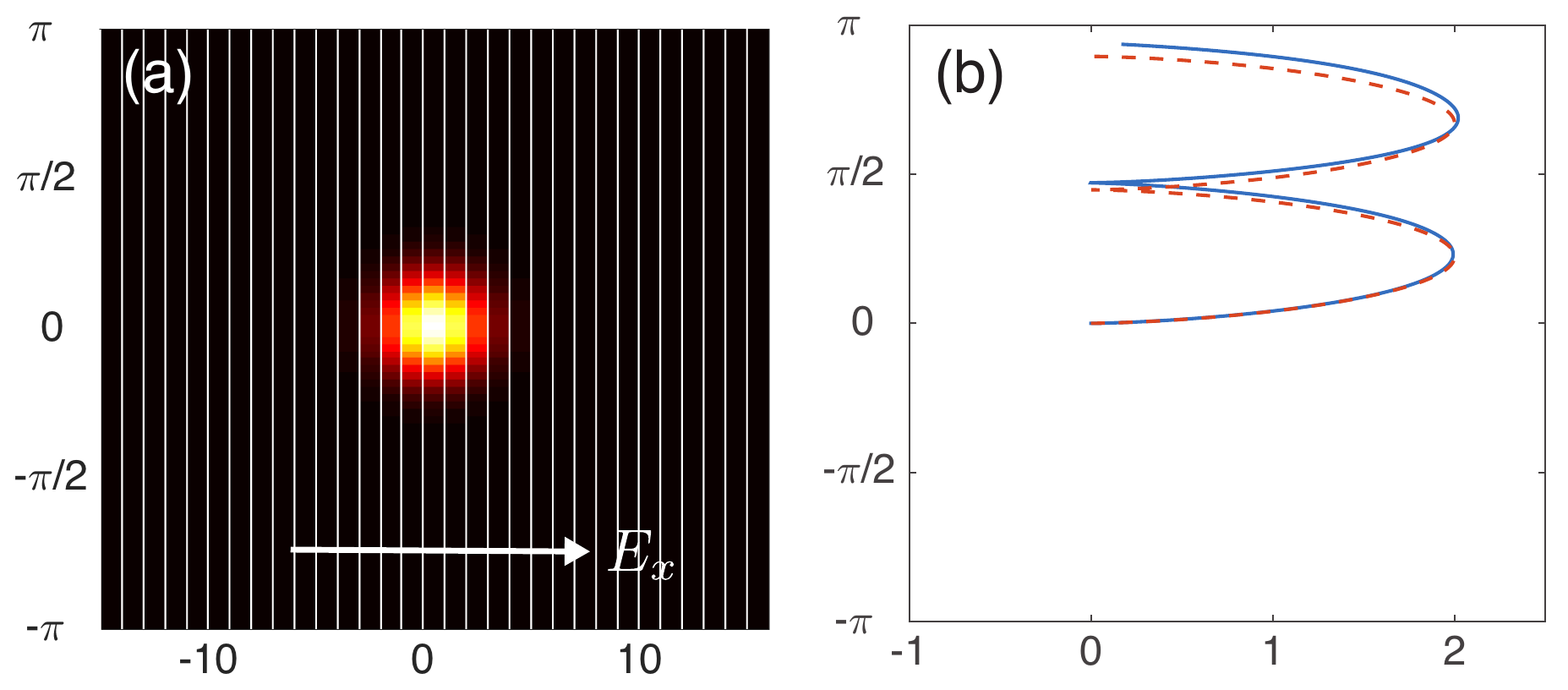}
\caption{Cyclotron motion of a wave packet for $N_x = 31$, $D = 0.1J$, $c = 1$, $UN = 20J$, and $E_x = 0.1J$, where $N$ is the total number of photons. In both (a) and (b), the horizontal axis is $x$ and the vertical axis is $\theta$. The initial condition of the wave packet is plotted in (a). In (b), the center-of-mass motion of the wave packet from $t = 0$ to $4\pi/\omega_c$ is plotted. The solid line is from the numerical simulation of the dynamics of the wave packet, and the dashed line is the analytical prediction from Eq.~(\ref{analyticalcom}).}
\label{cyclotron}
\end{center}
\end{figure}

In order to observe nontrivial orbits, we assume there is a constant force applied to the $x$ direction, which can be implemented via an additional small constant frequency gradient of cavities along the $x$ direction. Provided the initial wave packet is large enough to only explore the harmonic region of the spectrum at small $D$, the center of mass of the wave packet follows the semiclassical cyclotron orbit determined by the classical equations of motion:
\begin{align}
	m_x dv_x (t)/dt &= E_x + B_\mathrm{eff}v_\theta (t),\notag \\
	m_\theta d v_\theta (t)/dt &= -B_\mathrm{eff} v_x (t),
\end{align}
where $m_x = 1/2J$ and $m_\theta = 1/D$ are the effective masses in the $x$ and $\theta$ directions and $v_x(t)$ and $v_\theta (t)$ are the velocities of the center of mass at time $t$ in the $x$ and $\theta$ directions, respectively. The constant force $E_x$ is applied in the $x$ direction. Solving the equations under the initial conditions $x=v_x=v_\theta=0$ at $t=0$, the wave packet trajectory in time follows the curve
\begin{align}
	\langle x (t)\rangle &= (E_x/B_\mathrm{eff}^2)2 m_\theta\sin^2 \left( \omega_c t/2\right), \notag \\
	\langle \theta (t) \rangle &= (E_x/B_\mathrm{eff}^2)\sqrt{m_x m_\theta} \sin \left( \omega_c t\right) - (E_x/B_\mathrm{eff})t, \label{analyticalcom}
\end{align}
where $\omega_c \equiv B_\mathrm{eff}/\sqrt{m_x m_\theta}$ is the cyclotron frequency.

In Fig.~\ref{cyclotron}, we compare the analytical semiclassical prediction (\ref{analyticalcom}) for the center-of-mass motion of a wave packet obtained with a full numerical solution of the Hamiltonian (\ref{effham}) including interactions at the mean-field level. The agreement is good and, as expected, improves even further if smaller values of $D$ are used.

\begin{figure}[htbp]
\begin{center}
\includegraphics[width=8.0cm]{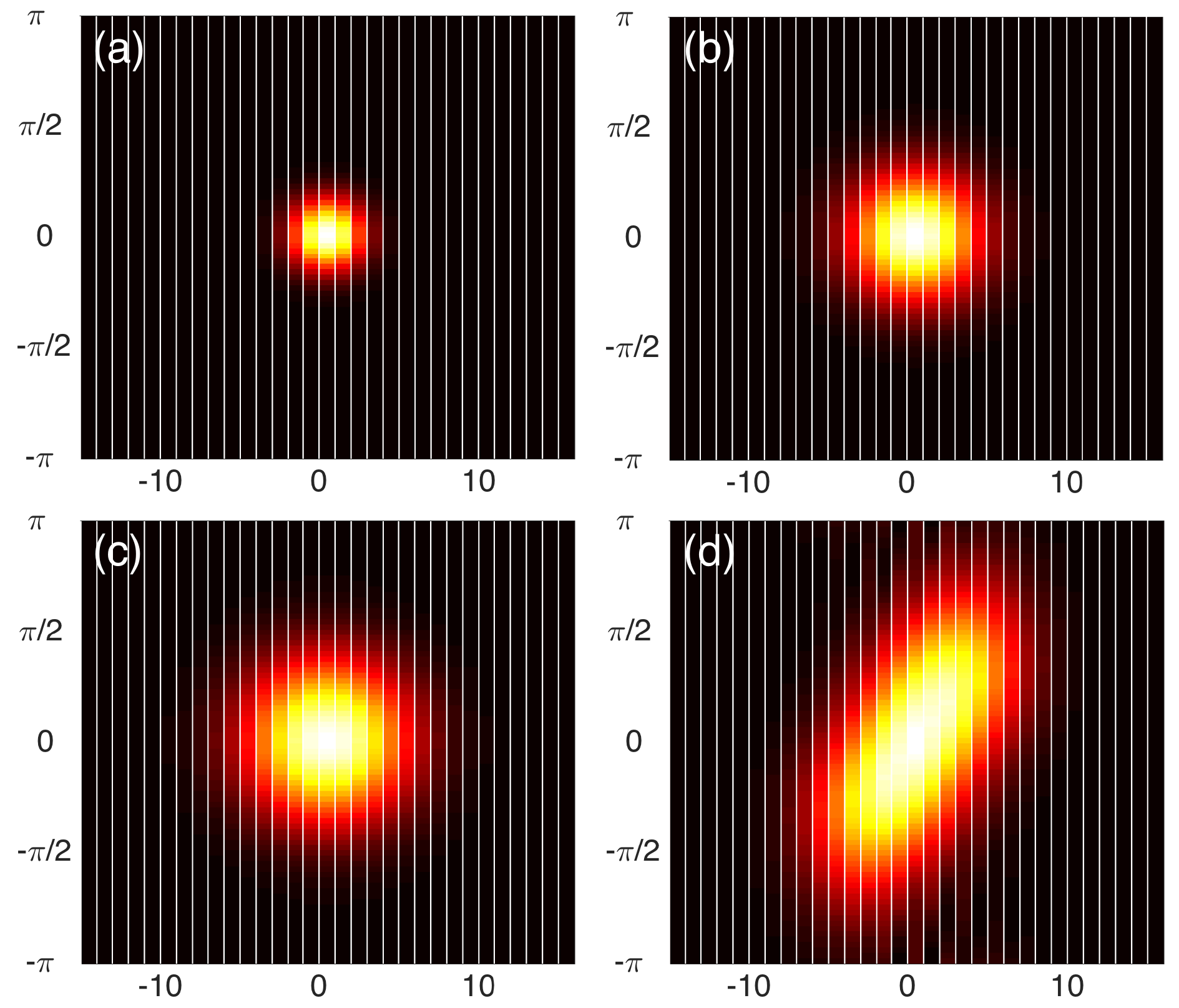}
\caption{Free expansion of a Gaussian wave packet for $c = 1$ and $D = 0.1J$ after the expansion time of $\pi/\omega_c$. (a) The initial wave packet. (b) The expansion in the absence of the interaction. (c) The expansion in the presence of the contact interaction with $UN = 20J$. (d) The expansion in the presence of the infinite-range interaction with $U_\mathrm{inf}N = 5J$. The horizontal axes are $x$, and the vertical axes are $\theta$.}
\label{exp_dyn}
\end{center}
\end{figure}

\textit{Expansion dynamics}.---
As a last point, we now present a scheme to experimentally assess the local nature of the photon-photon interactions.
We now assume there is no force ($E_x = 0$), and we treat the interactions at the mean-field level. 
In the absence of an effective magnetic field (i.e., $c = 0$), a circularly symmetric Gaussian wave packet in the $x$-$\theta$ plane maintains its circular symmetry during its expansion provided scaled axes are used so that the effective masses in the two directions are the same. (This can be achieved by introducing a scaled variable $y \equiv \theta \sqrt{2J / D}$ and preparing a symmetric wave packet in the $x$-$y$ plane.).
In the presence of a magnetic field ($c \neq 0$), the wave packet expansion is symmetric only when one has a rotationally symmetric gauge. Our Hamiltonian (\ref{effham}) is written in the Landau gauge where nontrivial hopping phases are present only along one direction. In order to observe a symmetric expansion, one needs to prepare a Gaussian wave packet with an appropriate phase, which corresponds to a gauge transformation from the Landau gauge to the symmetric gauge~\cite{Kim:1973,Tsuru:1992}.

In Fig.~\ref{exp_dyn}, we plot how such a suitably designed wave packet expands after half the period of the cyclotron motion in different cases. Fig.~\ref{exp_dyn}(a) shows the initial condition of the wave packet. Fig.~\ref{exp_dyn}(b) shows the expanded wave packet in the absence of interactions, when the ballistic expansion is due to diffraction. Fig.~\ref{exp_dyn}(c) shows the expanded wave packet for local interactions: interactions speed up the expansion, but the wave packet always maintains its initial symmetric shape as expected. 

For comparison, Fig.~\ref{exp_dyn}(d) shows the expansion for the infinite-ranged nonlocal interactions along $\theta$ of the form
$
	\mathcal{H}_\mathrm{inf}
	=
	\frac{U_\mathrm{inf}}{2} \sum_x
	\left( \int_0^{2\pi} d\theta b_x^\dagger (\theta) b_x (\theta) \right)^2,
$
which one would obtain, e.g., in the case of a temporally slow nonlinearity. In this case, the expansion along the $\theta$ direction does not initially feel the effect of the infinite-range interaction in the $\theta$ direction and is thus concentrated in the $x$ direction.
At later times, the elongated wave packet then rotates due to the synthetic Lorentz force induced by the synthetic magnetic field, which gives rise to the nonsymmetric final shape. Comparison of Figs.~\ref{exp_dyn}(c) and \ref{exp_dyn}(d) then shows a clear and experimentally observable signature of the local versus nonlocal interactions.

\textit{Outlook}.---
We have shown how local interactions can be combined with synthetic dimensions and strong synthetic magnetic fields in an array of nonlinear optical cavities. The exciting prospect of our scheme is to study four-dimensional interacting models in physically three-dimensional lattices. Once a medium with suitably strong photon-photon interactions~\cite{Schoelkopf:2008, You:2011, Gorshkov:2011,CarusottoCiuti} is included in the resonators, our proposal will open unprecedented possibilities to study fractional quantum Hall states and the other intriguing topological phases of matter in high dimensions~\cite{Zhang:2001,Kita}.

\begin{acknowledgments}
We are grateful to Hannah Price for stimulating discussions, and we also acknowledge valuable discussions with Nathan Goldman, Thomas Scaffidi, and Mohammad Hafezi. We thank Fernando Ramiro Manzano and Zeno Gaburro for useful exchanges on silicon microring resonators, and Albert Adiyatullin and Claud\'eric Ouellet-Plamondon for helpful discussions on their exciton-polariton micropillars.
This work was funded by ERC through the QGBE grant, by the EU-FET Proactive grant AQuS, Project No. 640800, and by the Provincia Autonoma di Trento, partially through the project ``On silicon chip quantum optics for quantum computing and secure communications -- SiQuro.''
\end{acknowledgments}

\end{document}


\title{Supplemental Material for ``Synthetic dimensions with magnetic fields and local interactions in photonic lattices"}

\author{Tomoki Ozawa}
\author{Iacopo Carusotto}
\affiliation{INO-CNR BEC Center and Dipartimento di Fisica, Universit\`a di Trento, I-38123 Povo, Italy}%

\date{\today}

\makeatletter
\renewcommand{\fnum@figure}[1]{FIG.~S\thefigure. }
\makeatother

\newcommand{\iac}[1]{{\color{red} #1}}
\newcommand{\tom}[1]{{\color{blue} #1}}

\maketitle
\appendix
\section{Effect of $w$-dependent hopping phases}

\begin{figure}[htbp]
\begin{center}
\includegraphics[width=8.5cm]{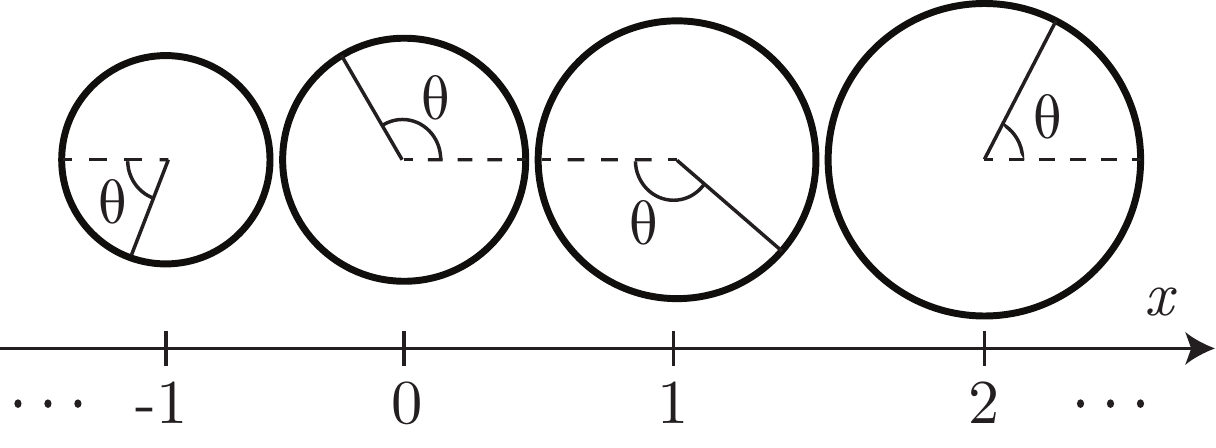}
\caption{The schematic illustration of hopping with alternating reference points for measuring the angle $\theta$. For each ring resonator, the dashed line indicates the reference direction from which the angle $\theta$ should be measured. }
\label{hop_SM}
\end{center}
\end{figure}

In this first part of the Sypplemental Material, we show that the hopping Hamiltonian Eq.~(5) in the Main Text can have a $w$-dependent hopping phase but this does not alter the main results of the paper.
For concreteness, we consider a situation where ring cavities are aligned and coupled along one direction. Taking into account the winding phase profile of ring cavity modes, and assuming that adjacent cavities are coupled along a short region, the hopping phase obtained as a photon hops from mode $w$ of the cavity located at $x$ to mode $w+c$ of the cavity located at $x+1$, is $e^{i(w + c)\pi}$.
The non-interacting tight-binding Hamiltonian analogous to Eq.~(5) taking into account this hopping phase is
\begin{align}
	\tilde{\mathcal{H}}_0 =&
	\sum_{x,w} \Omega_{x,w}a^\dagger_{x,w} a_{x,w} \notag \\
	&-
	\sum_{x,w} \left( J e^{i(w+c) \pi} a^\dagger_{x+1, w+c} a_{x,w} + \mathrm{H.c.}\right).
	\label{wdependent}
\end{align}
Going to the rotating frame by $b_{x,w} \equiv a_{x,w} \exp\{i [ \Omega_0 + \Omega_\mathrm{FSR} (w-w_0-xc)] t\}$ and moving to $x$-$\theta$ space by Fourier transformation, one arrives at
\begin{align}
	\mathcal{H}_0
	=&
	\sum_x \int_0^{2\pi}d\theta
	\left[
	\frac{D}{2}
	\{i\nabla_\theta b^\dagger_x (\theta)\}
	\{-i\nabla_\theta b_x (\theta)\}
	\right. \notag \\
	&\left.-J \left\{ e^{ic\theta + iw_0\pi}b^\dagger_{x+1}(\theta + \pi) b_x(\theta) + \mathrm{H.c.}\right\}
	\right].
\end{align}
Thus, comparing with the Hamiltonian (1) in the main text which we want to obtain, 
the hopping has acquired an additional constant phase  $e^{i w_0\pi}$, and the hopping from site $x$ to site $x+1$ is accompanied by the $\pi$ change of the angle $\theta$. It is now convenient to shift the origin of $\theta$ by $\pi$ for every other site, e.g. only for sites with odd $x$, which amounts to $b_{x}(\theta) \to b_x (\theta + \pi)$ for odd $x$. This shift of the origin of $\theta$ essentially corresponds to measuring $\theta$ with alternating reference point as indicated in Fig.~S\ref{hop_SM}.
The non-interacting Hamiltonian then becomes
\begin{align}
	\mathcal{H}_0
	=&
	\sum_x \int_0^{2\pi}d\theta
	\left[
	\frac{D}{2}
	\{i\nabla_\theta b^\dagger_x (\theta)\}
	\{-i\nabla_\theta b_x (\theta)\}
	\right. \notag \\
	&\left.-J \left\{ e^{ic\theta + i(w_0 + cx)\pi}b^\dagger_{x+1}(\theta) b_x(\theta) + \mathrm{H.c.}\right\}
	\right].
\end{align}
We have now recovered the non-interacting part of (1) in the main text up to additional hopping phases $e^{i(w_0 + cx)\pi}$. Such $\theta$-independent phases do not contribute to effective magnetic flux in $x$-$\theta$ plane and can be gauged-away, and therefore it does not alter the results of the paper.

\section{Edge properties of the model}

In this second part of the Supplemental Material, we complete the discussion in the main text by providing details on the edge properties of the Hamiltonian (1) of the Main Text in the absence of interactions. It is convenient to work in the Fourier transformed representation, Eq.~(3) in the main text, which reads
\begin{multline}
	\mathcal{H}=
	\\
	\sum_{x,w}\frac{D}{2}(w-w_0)^2 b^\dagger_{x,w}b_{x,w}
	-J
	\sum_{x,w} \left( b^\dagger_{x+1,w+c} b_{x,w} + \mathrm{H.c.} \right)
\end{multline}
The Hamiltonian can be made diagonal in $w$ by introducing a shifted operator
\begin{align}
	b_{x,w} = \tilde{b}_{x, w-cx}.
\end{align}
In terms of the shifted operator, the Hamiltonian becomes
\begin{align}
	\mathcal{H}
	=
	&\sum_{x,k_\theta}\frac{D}{2}(w - w_0)^2 \tilde{b}_{x,w - cx}^\dagger \tilde{b}_{x,w - cx}
	\notag \\
	&- J\left\{ \tilde{b}_{x+1,w-cx}^\dagger \tilde{b}_{x,w-cx} + \mathrm{H.c.} \right\}
	\notag \\
	=
	&\sum_{x,w}\frac{D}{2}(w - w_0 + cx)^2 \tilde{b}_{x,w}^\dagger \tilde{b}_{x,w}
	\notag \\
	&- J\left\{ \tilde{b}_{x+1,w}^\dagger \tilde{b}_{x,w} + \mathrm{H.c.} \right\}, \label{ham1d}
\end{align}
which is diagonal in $w$. By diagonalizing this Hamiltonian for a given value of $w$ with an open boundary condition in $x$ direction, we obtain the energy spectrum of the system in the presence of edges in $x$ direction. One can observe that the first term acts as a harmonic trapping potential in $x$ direction, centered at $-(w - w_0)/c$; this structure is very similar to the case of an ordinary two dimensional Landau level problem in the Landau gauge. As in the Landau level problem, the effect of the finite size of the system is felt when the center of the trapping potential $-(w - w_0)/c$ reaches the edge.

\begin{figure}[htbp]
\begin{center}
\includegraphics[width=8.5cm]{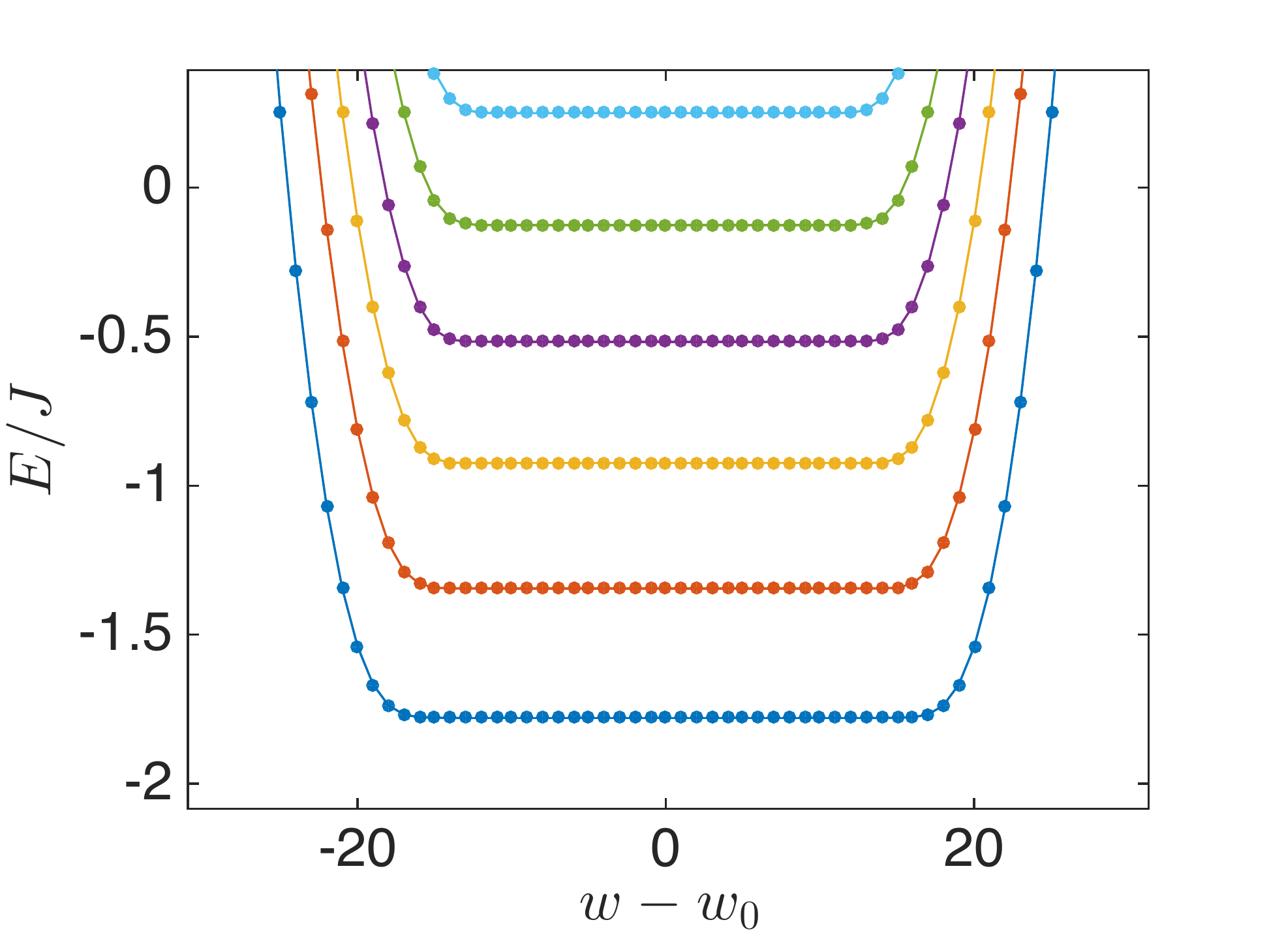}
\caption{The energy spectrum as a function of $w - w_0$ under the open boundary condition for $N_x = 41$. We chose $D = 0.1J$ and $c = 1$. Since $w - w_0$ can take only integer values, the lines connecting the dots are just a guide to the eye.}
\label{edgespectrum}
\end{center}
\end{figure}

In Fig.~S\ref{edgespectrum}, we plot the spectrum as a function of $w - w_0$, calculated for a system with $N_x = 41$ sites in $x$ direction with $c = 1$, and we chose $-20 \le x \le 20$. The spectrum is reminiscent of the Landau level problem.
There are flat levels for small values of $|w - w_0|$, which are bulk degenerate levels of the system. In the region where $|w - w_0|$ is large and of order $cN_x/2$, the harmonic oscillator center in (\ref{ham1d}) reaches the edge of the systems, and the energy levels are not flat anymore. We find that the wavefunctions satisfying $w - w_0 \gtrsim cN_x/2$ are localized at the right edge of the array, whereas the wavefunctions satisfying $w - w_0 \lesssim -cN_x/2$ are localized at the left edge. The states with $w - w_0 \gtrsim cN_x/2$ localized at the right edge have a positive group velocity, thereby forming edge states which move only in the positive $\theta$ direction. On the other hand, the states with $w - w_0 \lesssim -cN_x/2$ localized at the left edge have a negative group velocity, forming edge states which move only in the negative $\theta$ direction. This unidirectionality of states localized at the edge indicates that we have chiral edge states, which are characteristic of two dimensional topological models with time-reversal symmetry breaking.

